\documentclass[aps,twocolumn,pre,showpacs]{revtex4-1}

\usepackage{amsmath}%
\usepackage{amssymb}%
\usepackage{graphicx}% Include figure files
\usepackage{dcolumn}% Align table columns on decimal point
\usepackage{bm}% bold math
\usepackage{stmaryrd}%
\usepackage[shortlabels]{enumitem}%

\newcommand{\bbE}{\mathbb{E}}%
\newcommand{\Z}{\mathbb{Z}}%

\begin{document}

\title{Subthreshold behavior and avalanches in an exactly solvable
  Charge Density Wave system}%

\author{David~C.~Kaspar$^{1}$\thanks{ {\tt kaspar@math.berkeley.edu}} and Muhittin Mungan$^2$\thanks{ {\tt  mmungan@boun.edu.tr}} } 
\affiliation{$^1$  Mathematics Department, University of California,   Berkeley, CA 94720, USA,}
\affiliation{$^2$  Physics Department,   Bo\u{g}azi\c{c}i University,   Bebek 34342 Istanbul, Turkey }

\date{\today}


\begin{abstract} 
 
We present a toy charge density wave (CDW) model in 1d
  exhibiting a depinning transition with threshold force and
  configurations that are explicit.  Due to the periodic boundary
  conditions imposed, the threshold configuration has a set of
  topological defects whose location and number depend on the
  realization of the random phases.
  Approaching threshold, these defects are relocated by avalanches
  whose size dependence on the external driving force $F$ is described
  by a record-breaking process. We find that the depinning transition
  in this model is a critical phenomenon, with the cumulative
  avalanche size diverging near threshold as $(F_{\rm th} -
  F)^{-2}$. The exact avalanche size distributions and their
  dependence on the control parameter $(F_{\rm th} - F)$ are obtained.
  Remarkably, the scaling exponents associated with the critical
  behavior depend on (1) the initial conditions and (2) the
  relationship between the system size and the pinning strength.
\end{abstract}

\maketitle

The motion of Charge Density Waves (CDW) belongs to the class of
systems in which an elastic structure is driven by external forces
through a random medium.  Fisher \cite{DSFisher83,DSFisher85} has
argued that the depinning transition, when the deformable medium
begins to slide, is a \emph{dynamic critical phenomenon}: a phase
transition with the driving force as the control parameter.
Analytical results for the divergence of strains \cite{snc90,snc91},
functional renormalization group calculations \cite{NarayanFisher92,
  LeDoussal02,Ertas}, and extensive numerical simulation of CDW and
similar systems in dimensions $d = 1,2,3$ \cite{Littlewood86,
  Veermans90, MyersSethna93, Rosso02, Jensen, Middleton93,
  NarayanMiddleton94} support this claim.  However, there are few
\emph{rigorous} results providing evidence whether the depinning
transition is indeed a critical phenomenon, particularly in $d = 1$.
In this letter we introduce an exactly solvable CDW model in $d = 1$
that exhibits the tell-tales of a critical phenomenon and allows us to
understand the origin of criticality.

We begin with the CDW Hamiltonian and some accompanying notation.  Let
\begin{equation}
  \mathcal{H}(\{y_i\}) = 
  \sum_i \frac{1}{2}(y_i - y_{i-1})^2 + V(y_i - \alpha_i) - Fy_i,
  \label{eqn:Ham}  
\end{equation}
where $V(x)$ is 1-periodic and $\alpha_i$, the impurity phases, are
\emph{i.i.d.}\ uniform on $(-\frac{1}{2},+\frac{1}{2})$. Following
Narayan and Fisher \cite{NarayanFisher92}, we choose
\begin{equation}
  V(x) = \frac{\lambda}{2} (x - \llbracket x \rrbracket)^2.
  \label{eqn:Pot}  
\end{equation}
Here $\lambda$ is the strength of the potential and $\llbracket x
\rrbracket$ is the nearest integer to $x$. Write
\begin{equation}
  m_i \equiv \llbracket y_i - \alpha_i \rrbracket 
  \qquad \text{and} \qquad 
  \tilde{y}_i \equiv y_i - \alpha_i - m_i, 
\end{equation}
for the \emph{well number} and \emph{well coordinate} of $y_i$; the
former records which parabolic well contains $y_i$ and the latter the
displacement in $(-\frac{1}{2}, +\frac{1}{2}]$ from the well's center.

For static configurations the piecewise-parabolic potential permits an
explicit formula for $\tilde{y}_i$ in terms of the well numbers $m_i$
and the phases $\alpha_i$,
\begin{equation}
  \tilde{y}_i = \frac{\eta}{1-\eta^2} 
  \sum_{j\in\mathbb{Z}} \eta^{|i - j|} (\Delta \alpha_j + \Delta m_j) + F/\lambda.
  \label{eqn:tildey}
\end{equation}
Here $\Delta$ is the discrete Laplace operator, $\Delta \alpha_i =
\alpha_{i-1} - 2\alpha_i + \alpha_{i+1}$, and $0 < \eta < 1$ is the
smaller root of $\eta^2 - (2 + \lambda) \eta + 1 = 0$.  The
nonlinearity of the system is an admissibility condition: $m_i$ must
be such that \eqref{eqn:tildey} gives $\tilde{y}_i \in (-\frac{1}{2},
+\frac{1}{2}]$.  One can check that all static configurations are
linearly stable unless a particle is at a cusp of $V$.  We investigate
the case where $m_i$ and $\alpha_i$ are $L$-periodic, the latter still
\emph{i.i.d.}\ within a single period.

Consider the behavior of this system starting from some initial
configuration and increasing the force.  We assume that the dynamics
of the particles are purely relaxational and that all changes in the
force are sufficiently slow that we reach the static configurations.
From \eqref{eqn:tildey} it follows that increasing $F$ translates all
particles uniformly until the first particle, say $j$, reaches a cusp,
$\tilde{y}_j = 1/2$.  Any further infinitesimal increase causes
particle $j$ to jump wells $m_j \rightarrow m_j + 1$, displacing
\begin{equation}
  \tilde{y}_k \rightarrow \tilde{y}_k - \delta_{jk} +
  \frac{1-\eta}{1+\eta} \;  \eta^{\vert j - k \vert}, \label{eqn:tildeyresp}
\end{equation}
which may cause other particles to cross $+\frac{1}{2}$ and jump as
well, and so on.  Depending on $F$ this process may terminate,
yielding another stable configuration, or continue forever, which we
interpret as the sliding state.

Like the CDW models with sinusoidal potential \cite{Middleton93,
  NarayanMiddleton94}, this model exhibits behavior which is:
\begin{itemize}
\item \emph{Irreversible.}  If we change the force by $\Delta F$,
  causing one or more particles to jump wells before reaching another
  stable configuration, and then return $F$ to its original value,
  \emph{none} of the particles jump back \cite{KM12}.
\item \emph{Reversible.}  The resulting configuration reacts by rigid
  translation, without jumps, to forces in the interval between $F$
  and $F + \Delta F$.
\end{itemize}
To focus on the irreversible dynamics, we identify configurations at
nonzero $F$ with their $F = 0$ versions (\emph{zero-force
  configurations}) which have the same well numbers $m_i$, provided
that the well coordinates $\tilde{y}_i$ of the latter react to $F$ by
rigid translation only.

\section{The toy model}
We simplify our CDW model further by choosing $\lambda$ large ($\eta$
small): defining rescaled well coordinates $z_i$ by $\eta z_i =
\tilde{y}_i - F/\lambda$, we obtain from \eqref{eqn:tildey}
\begin{equation}
  z_i = \Delta \alpha_i + \Delta m_i + O(\eta).
  \label{eqn:zeq}
\end{equation}
We will refer to the CDW model, dropping $O(\eta^2)$ terms ($O(\eta)$
after rescaling), as the \emph{toy model}; here we have
nearest-neighbor interactions only and obtain a set of exact results.
Proofs and additional results for the untruncated model will be given
elsewhere \cite{KM12}.

For the toy model, the process of increasing the external force and
evolving the configurations to threshold can be described by the
following \emph{zero-force algorithm} (ZFA), which operates on
$L$-periodic zero-force configurations:
\begin{itemize}
\item[] (ZFA1) Record $z_{\rm max} = \max_i z_i$.
\item[] (ZFA2) Find $j = {\rm arg} \,{\rm max}_i z_i$ and set
  \begin{equation} \label{eqn:ZFA}
    \begin{aligned}
      m_j &\to m_j + 1 \\
      z_{j\pm 1} &\to z_{j\pm 1} + 1 \\
      z_j &\to z_j - 2.
    \end{aligned}
  \end{equation}
\item[] (ZFA3) If any $z_i > z_{\rm max}$ from (ZFA1), goto (ZFA2).
\end{itemize}
The initial execution of (ZFA2) jumps the particle which would first
reach the cusp \emph{if} the force were increased, and subsequent
executions (if any) resolve those particles which would be pulled
across the cusp as a result of the first.  We note the following
properties of the ZFA:
\begin{itemize}
\item[(i)] The ZFA always terminates, with all sites having jumped at
  most once.
\item[(ii)] If in secondary executions of (ZFA2) we take $j$ to be any
  site where $z_j > z_{\rm max}$, not necessarily the maximum, the
  final result is unchanged.
\item[(iii)] If all sites jump, then the $z_i$ are unchanged, and this
  fixed-point is the threshold configuration.
\item[(iv)] The ZFA finds the threshold configuration after a number
  of iterations which is bounded by a function of the system size $L$.
\end{itemize}

As noted for similar CDW models \cite{Tangetal87, MyersSethna93,
  NarayanMiddleton94}, (ZFA2) suggests a connection to Abelian
sandpile models; see Redig \cite{Redig05} for an introduction.  In
that setting, the change \eqref{eqn:ZFA} in $z$ would be called
\emph{toppling} at $j$, and (ii) above is precisely the Abelian
property.  Note, however, that although a variety of sandpile models
with slightly varying features have been studied previously (see
\cite{Turcotte99} for a survey and \cite{Zhang89} for a model which
also has continuous heights), what we call the toy model does not seem
to be among them.  Our model has periodic boundary; is conservative in
that $z$ ``mass'' is moved about, but neither added nor removed;
evolves deterministically, with integer changes only; and has a random
fractional part from the initial conditions which is preserved by the
dynamics.

\section{Threshold configurations}
The ZFA indicates that the evolution to the depinning transition under
force increments minimizes $\max_i z_i$, and indeed the threshold
configuration is the solution of the variational problem\footnote{The
  same is in fact true for the untruncated model with the
  corresponding adaptation of the ZFA \cite{KM12}.} $\min_m \max_i
z_i$.  From this perspective \eqref{eqn:zeq} suggests $\Delta m_i =
-\llbracket \Delta \alpha_i\rrbracket$ would be favorable, but
periodicity and the requirement that $m_i \in \Z$ usually prevent
this.  We need only choose to deviate from this guess in the most
favorable places.

Let $S \equiv \sum_{i=0}^{L-1} \llbracket \Delta \alpha_i \rrbracket$.
The threshold configuration $\{m^+_i\}$ satisfies
\begin{equation}
  \Delta m^+_i = 
  - \llbracket \Delta \alpha_i \rrbracket + J^+_i - \delta_{i k^+}
  \label{eqn:Deltam}
\end{equation}
where $J$ is an integer vector selected as follows:
\begin{itemize}
\item[(i)] \emph{Case $S \geq 0$.} $J^+_i = 1$ for the $S + 1$
  positions $i$ which have smallest $\Delta \alpha_i - \llbracket
  \Delta \alpha_i \rrbracket$, $J^+_i = 0$ otherwise;
\item[(ii)] \emph{Case $S < 0$.} $J^+_i = -1$ for the $|S| - 1$
  positions $i$ which have largest $ \Delta \alpha_i - \llbracket
  \Delta \alpha_i \rrbracket$, $J^+_i = 0$ otherwise;
\end{itemize}
and $k^+$ is an index defined by (\emph{divisibility condition})
\begin{equation}
  k^+ \equiv 
  \sum_{i=0}^{L-1} i (-\llbracket \Delta \alpha_i \rrbracket + J^+_i) \pmod{L}.
  \label{eqn:div}
\end{equation}
The proof is straightforward and given in \cite{KM12}.  We will refer
to those sites where $\Delta m^+_i + \llbracket \Delta \alpha_i
\rrbracket \equiv \epsilon_i \neq 0$ as \emph{defects} with
\emph{charge} $\epsilon_i$.

\section{The depinning threshold force}
Using \eqref{eqn:Deltam}, the threshold configuration $z_i^+$ for the
toy model is explicit, as is the threshold force:
\begin{equation}
  F_{\rm th}\left ( \{\alpha_i\} \right ) = 
  \lambda \left ( 1/2 - \eta z^+_{\rm max} \right ),
  \label{eqn:fth}
\end{equation}
where $z^+_{\rm max} \equiv \max_i z^+_i$.  The term in parentheses on
the right-hand side is the distance from the particle with maximum
$z_i$ to the cusp.  Defining
\begin{equation}
  \omega_i \equiv \Delta \alpha_i -
  \llbracket \Delta \alpha_i \rrbracket
\end{equation}
and using \eqref{eqn:zeq} and \eqref{eqn:Deltam}, we find
that\footnote{with probability $1-L^{-1}$ \cite{KM12}}
\begin{equation}
  z_{\rm max}^+   
  = \begin{cases}
    \omega_{\sigma(S)} + 1 & \text{if } S \geq 0 \\
    \omega_{\sigma(L-|S|)} & \text{if } S < 0
  \end{cases}
\end{equation}
where $\sigma$ is the permutation of the indices that orders $\omega$:
\begin{equation}
  \omega_{\sigma(0)} < \omega_{\sigma(1)} < \cdots <
  \omega_{\sigma(L-1)}. \nonumber
\end{equation}
We therefore understand the dependence of $z^+_{\rm max}$ on the
random phases $\alpha_i$ as coming almost exclusively from the sum $S
= -\sum_{i=0}^{L-1} \omega_i$ and rank statistics of $\{\omega_i\}$.

It turns out \cite{KM12} that any $L-1$ (but not all $L$) of the
$\omega_i$ are \emph{i.i.d.}\ uniform $(-\frac{1}{2},+\frac{1}{2})$,
whence routine arguments show $L^{-1/2} S$ converges in distribution
as $L \to \infty$ to a normal random variable with mean $0$ and
variance $1/12$.  Thus the typical number of topological defects
scales as $L^{1/2}$, and the threshold force averaged over the
quenched disorder is found to be
\begin{equation}
  \bbE F_{\rm th} = (1-\eta)^3/2\eta + O(L^{-3/2}).
\end{equation}
The variance $(\Delta F_{\rm th})^2$ scales as
\begin{equation}
  (\Delta F_{\rm th})^2 \sim L^{-1}.
\end{equation}
so that fluctuations $\Delta F_{\rm th}$ scale as $L^{-1/2}$.  This
matches the scaling behavior of the sinusoidal CDW model
\cite{Middleton93,NarayanMiddleton94}, for which the expected behavior
is $\Delta F_{\rm th} \sim L^{-1/\nu_{FS}}$, with $\nu_{FS}$ the
finite-size scaling exponent from the scaling theory of Chayes {\em et
  al} \cite{Chayes86}.  Our model saturates their prediction $\nu_{FS}
\ge 2/d$ in $d = 1$ with
\begin{equation}
  \nu_{FS} = 2.
\end{equation}

\section{Flat IC to threshold}
We are interested not only in the threshold state itself, but also the
changes that occur as we drive the system toward it.  One might start
with any stable configuration, but two initial conditions seem
particularly natural.  Here we treat the ``flat'' case, $m_i = 0$ for
all $i$, and address the other in the next section.

The primary quantity of interest is the \emph{cumulative avalanche
  size} $\Sigma$, \emph{i.e.}~the total number of jumps which occur,
or equivalently the \emph{polarization} $P = \Sigma/L$.  For the flat
IC,
\begin{equation}
  \Sigma = \sum_{i=0}^{L-1} m^+_i
  \label{eqn:flatSigma}
\end{equation}
where $\{m^+_i\}$ are the well numbers of the unique threshold
configuration with $\min_i m^+_i = 0$.  The characterization
\eqref{eqn:Deltam} can be used to obtain the scaling behavior of
$\Sigma$ and $P$ with $L$.

The key observation is that components of $\{z^+_i\}$ are
\emph{exchangeable}, i.e.~their joint distribution is invariant under
permutations \cite{KM12}.  This leads directly to a scaling limit for
the threshold \emph{strains} $s_i \equiv m^+_i - m^+_{i-1}$.  Define a
\emph{rescaled strain process} $s^{(L)}(t)$ by
\begin{equation}
  s^{(L)}(t) \equiv (L/12)^{-1/2} s_{\lfloor Lt \rfloor} \quad (0 \leq t
  \leq 1).
  \label{eqn:restrain}
\end{equation}
These processes in the \emph{Skorokhod space}\footnote{consisting of
  functions which are right-continuous with left-limits}
$\mathcal{D}([0,1])$ converge in distribution to a Brownian bridge
with zero integral \cite{KM12}.  More precisely, the limiting
distribution is that of
\begin{equation}
  B(t) - \int_0^1 B(r) \, {\rm d} r
\end{equation}
where $B(t)$ is the Brownian bridge, the result of conditioning
Brownian motion to return to $0$ at $t = 1$.

This implies that the typical maximum and minimum of $(s_i)$ at
threshold is diverging like $L^{1/2}$, as expected from the work of
Coppersmith \cite{snc90,snc91}, but also more: as $(s_i)$ is the
discrete derivative of $(m^+_i)$, we integrate it twice \cite{KM12} to
find that
\begin{equation}
  \Sigma \sim L^{5/2} \quad \text{and thus} \quad P \sim L^{3/2}
  \label{eqn:flatscaling}
\end{equation}
for typical realizations of the randomness.

\section{Threshold to threshold}
Taking the \emph{negative} threshold configuration as our initial
condition \cite{Middleton93, NarayanMiddleton94}, we can give a more
complete picture, including the beginning, the end, and also the
intermediate states observed as we iterate the ZFA.

We first adapt \eqref{eqn:Deltam} for the negative
threshold configuration, which maximizes $\min_i z_i$.  Define $J^-$
and $k^-$:
\begin{itemize}
\item[(i)] \emph{Case $S > 0$.} $J^-_i = 1$ for the $S - 1$ positions
  $i$ which have smallest $\omega_i$, $J^-_i = 0$
  otherwise; %\label{eqn:Sge0}
\item[(ii)] \emph{Case $S \leq 0$.} $J^-_i = -1$ for the $|S| + 1$
  positions $i$ which have largest $\omega_i$, $J^-_i = 0$
  otherwise; %\label{eqn:Slt0}
\end{itemize}
and $k^-$ is given in terms of $J^-$ by analogy with \eqref{eqn:div}.

It is convenient to introduce
\begin{equation}
  \zeta_i = \omega_i + J^-_i
\end{equation}
and the permutation $\pi$ that orders $\zeta$:
\begin{equation}
  \zeta_{\pi(0)} < \zeta_{\pi(1)} < \cdots <
  \zeta_{\pi(L-1)}.
  \label{eqn:zeta_order}
\end{equation}
Note $\zeta_{\pi(L-1)} - \zeta_{\pi(0)} < 1$. The $(\pm)$-threshold
configurations have
\begin{align}
  z^-_i &= \zeta_i + \delta_{ik^-} \label{eqn:zminus} \\
  z^+_i &= \zeta_i + \delta_{i \pi(0)} + \delta_{i \pi(1)} -
  \delta_{ik^+}, \label{eqn:zplus}
\end{align}
and, using the divisibility condition, $k^\pm$ are related
by\footnote{Here and in the the following addition and subtraction of
  indices are mod $L$.}
\begin{equation}
  k^+ = \pi(0) + \pi(1) - k^-.
  \label{eqn:kplus}
\end{equation}

Applying the ZFA to the negative threshold configuration, 
$z_{\rm max} = \zeta_{k^-} + 1$ and $k^-$ is the first site to jump:
\begin{align*}
  z_{k^-} &= \zeta_{k^-} + 1 & \to &&
  z_{k^-} &= \zeta_{k^-} - 1 \\
  z_{k^- \pm 1} &= \zeta_{k^- \pm 1} & \to && z_{k^- \pm 1} &=
  \zeta_{k^- \pm 1} + 1.
\end{align*}
The neighboring sites $k^- \pm 1$ will be forced to jump
if $\zeta_{{k^-} \pm 1} > \zeta_{k^-}$; this process continues for
all those \emph{consecutive} neighbors $i$ to the left and right of
$k^-$ for which $\zeta_i > \zeta_{k^-}$.  Write $i_L$ and $i_R$ for
the first sites on the left and right, respectively, of $k^-$ which
have
\begin{equation}\label{eqn:omegarec}
  \zeta_{i_L}, \zeta_{i_R} \leq \zeta_{k^-}
\end{equation}
and thus will \emph{not} be forced to jump.

If jumps occur at both sites $k^- \pm 1$, then $z_{k^-}$ is unchanged,
hence still the maximum, and the next ZFA also begins at $k^-$.  This
observation allows us to identify the \emph{avalanche} consisting of
all consecutive ZFA iterations initiated at a common site; the
individual iterations which constitute it will be called
\emph{avalanche waves} \cite{Priezzhev94}.  Given an initial site $i$,
with sites $i_L$ and $i_R$ --- the closest on the left and right which
do not jump in the first avalanche wave --- the avalanche finishes
after $\min(i-i_L, i_R-i)$ waves with a total of $(i-i_L)(i_R-i)$
jumps \cite{KM12}.

To better understand the threshold-to-threshold evolution under the
ZFA, observe that the ranks $\pi(j)$ of the $\zeta_j$ suffice to
determine the avalanche's initial site and extents.
We represent a given configuration $z_j$ by displaying
the rank $\pi(j)$ of $\zeta_j$ and using over- or underlines to
indicate additions by $\pm 1$ which are acquired as a result of jumps:
\begin{eqnarray}
  \overline{\pi(j)}  &\leftrightarrow&  z_{\pi(j)} = \zeta_{\pi(j)} + 1,  \\
  \underline{\pi(j)}  &\leftrightarrow&  z_{\pi(j)} = \zeta_{\pi(j)} - 1. 
\end{eqnarray}

For example, suppose that $z^-$ has the rank representation
\begin{equation*}
  \begin{array}{cccccccccccc}
    \ldots & 0 & 10 & 12 & 17 & \overline{15} & 16 & 18 & 11 & 13 & 1 & \ldots    
  \end{array}
\end{equation*}
so that $\pi(k^-) = 15$.  The extents of the first avalanche wave are
$k^- - i_L = 2, i_R - k^- = 3$ and after the sites bracketed below
have jumped, the resulting configuration is
\begin{equation*}
  \begin{array}{cccccccccccc}
    \ldots & 0 & 10 & \overline{12} & [\underline{17} &  \overline{15} 
    & 16 & \underline{18}] & \overline{11} & 13 & 1 & \ldots.    
  \end{array}
\end{equation*}
In the second wave, $k^-$ and $k^- + 1$ jump again, yielding
\begin{equation*}
  \begin{array}{cccccccccccc}
    \ldots & 0 & 10 & \overline{12} & 17 & [15 
    & \underline{16}] & 18 & \overline{11} & 13 & 1 & \ldots,    
  \end{array}
\end{equation*}
and the avalanche is complete.  The remaining avalanches begin at the
sites ranked $12$, $11$, and $10$; the result is the positive
threshold configuration.

This example illustrates that the important sites in the
threshold-to-threshold evolution are the 
\emph{lower records}\cite{Glick,RecordsBook}: given a
sequence of values $X_1, X_2, \ldots$, we say that $X_i$ is a
\emph{lower record} if $X_i = \min\{X_j : j \leq i\}$.  Using
\eqref{eqn:omegarec} we see that avalanches are determined by the
locations of the lower records of the sequences
\begin{align}
  \mathcal{J}_L &=
  \zeta_{k^-}, \zeta_{k^- - 1}, \zeta_{k^- - 2}, \ldots, \zeta_{\pi_L},
  \label{eqn:sequencesL} \\
  \text{and} \quad \mathcal{J}_R &= \zeta_{k^-}, \zeta_{k^- + 1},
  \zeta_{k^- + 2}, \ldots, \zeta_{\pi_R},
  \label{eqn:sequencesR}
\end{align}
where $\{\pi_L,\pi_R\} = \{ \pi(0), \pi(1) \}$ are the
termination sites. The evolution from negative to positive threshold
terminates when the avalanches reach $\pi_L$ and $\pi_R$.

For use in the next section we note the following \cite{KM12}:
\begin{itemize}
\item The variates $\zeta_i$ are exchangeable.
\item Thus $\pi_L$ and $\pi_R$ are selected uniformly from all
  pairs of (distinct) indices.
\item In fact, $k^-$ is independent of $\pi$.
\end{itemize}

\section{Avalanche size distributions}
We are interested in the cumulative
avalanche size $\Sigma$ and obtain in this section a parametric
family of distributions in the $L \to \infty$ limit.

As we move from $(-)$-threshold to $(+)$-threshold, let $X = z_{\rm
  max} - z^+_{\rm max}$ be the current maximum height minus that of
the $(+)$-threshold configuration; $X \in [0,1)$.  Shift indices so
that $k^- = 0$, and let $j_L(x)$ and $j_R(x)$ be the noninclusive left
and right endpoints of the interval of sites that have jumped in the
threshold-to-threshold evolution to achieve $X \leq x$, chosen so that
$-L < j_L \leq 0 \leq j_R < j_L + L$.  Observe that $j_L$ and $j_R$
take values which are indices of lower records: in fact, $j_L(x)$ and
$j_R(x)$ are the first indices in $\mathcal{J}_L$ and $\mathcal{J}_R$
for which $\zeta_i \leq \zeta_{\pi(1)} + x$.

If $\{m_i(x)\}$ are the well numbers of the first configuration with $X
\leq x$, the corresponding avalanche size is
\begin{equation}
  \Sigma(x) = \sum_{i=0}^{L-1} (m_i(x) - m^-_i) = |j_L(x)| j_R(x). \label{eqn:SigmajLR}
\end{equation}
The simplicity of \eqref{eqn:SigmajLR} is due to the
geometry of $\{m_i(x) - m^-_i\}$ in the threshold-to-threshold case,
which deviates from $0$ in only a single (discrete) trapezoidal bump
\cite{KM12}.  Through \eqref{eqn:SigmajLR} we address the statistics
of $\Sigma(x)$.

For large $L$, approximate $\zeta_{\sigma(1)} = -\frac{1}{2}$ and
$\zeta$ in $\mathcal{J}_L$ and $\mathcal{J}_R$ as \emph{i.i.d.}\
uniform $(-\frac{1}{2},+\frac{1}{2})$ variates, sharing their first
elements.  Then $j_L$ and $j_R$ are truncated\footnote{The system size
  $L$ bounds $j_L + j_R$.} geometric random variables.  Setting $u =
Lx$, one shows shows that as $L \to \infty$ the distribution of the
pair
\begin{equation}
  \left(\frac{|j_L(u/L)|}{L}, \frac{j_R(u/L)}{L}\right)
\end{equation}
converges to a truncation of a pair of exponential random variables
with rate $u$.  One then obtains for each value of $u$ the limiting
probability density of the rescaled avalanche size, $\varsigma(u)
\equiv \lim_{L \to \infty} \Sigma(u/L)/L^2$ \cite{KM12}:
\begin{equation}
  p_u(s) = 
  \int_{2\sqrt{s}}^1 {\rm d} z \, e^{-zu} \frac{4+8u(1 - z)+ 2u^2(1-z)^2}{(z^2 - 4s)^{1/2}},
  \label{eqn:pdist}
\end{equation}
which is supported on $[0,\frac{1}{4}]$.  In particular, when $u = 0$,
we obtain the distribution of the rescaled total number of jumps
$\varsigma(0)$ in the threshold-to-threshold evolution,
\begin{equation}
  p_0(s) = 2 \ln \frac{1 + \sqrt{1-4s}}{1 - \sqrt{1-4s}}.
  \label{eqn:PSigmath2th} 
\end{equation}
Note that \eqref{eqn:PSigmath2th} is precisely the avalanche size
distribution of Dhar's Abelian sandpile model in $d = 1$ considered by
Ruelle and Sen \cite{Redig05,Dhar,RS92,Pruessner}. This connection is
not an accident \cite{KM12}: the total threshold-to-threshold
evolution, without intermediate details, maps to a recurrent sandpile
configuration, a site at which sand is added, and the resulting
toppling.  However, this map loses information, specifically the
ordering and values of the well coordinates $z_i$, which are necessary
to obtain the complete family \eqref{eqn:pdist}.

\begin{figure}[t!]
  \includegraphics[scale = 0.3]{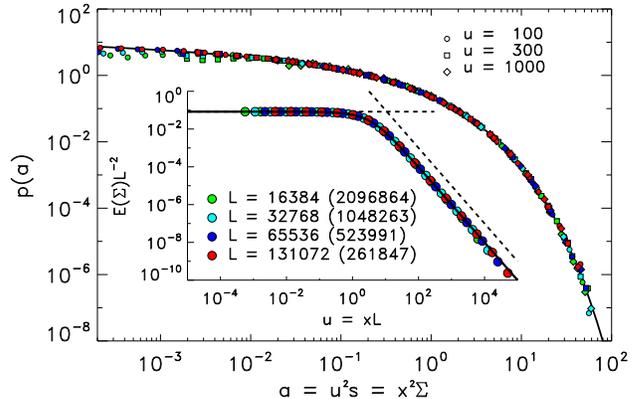}
  \caption{Numerical cumulative avalanche size distribution for
    various $L$ and $u$.  For large $u$, the distributions collapse
    when avalanche sizes are scaled as $a = u^2 s$. The solid line is
    \eqref{eqn:pdist}. Inset: Finite-size scaling behavior of the mean
    cumulative avalanche size.  The horizontal and slanted dashed line
    correspond to $\Phi(0)$ in \eqref{eqn:Phizero}, and a power law of
    exponent $-2$, respectively.  The solid line is the finite size
    scaling function \eqref{eqn:Phi}.  Symbol colors refer to
    different $L$ as indicated in the legend, where the numbers of
    realizations are shown in parentheses.}
  \label{polfig}
\end{figure}

From \eqref{eqn:pdist} we find the rescaled mean avalanche size,
$\lim_{L\to\infty} L^{-2} \mathbb{E}[\Sigma(u/L)] \equiv \Phi(u)$, as
\begin{equation}
  \Phi(u) = \frac{6 - 4u + u^2 - 6e^{-u} - 2ue^{-u}}{u^4} \label{eqn:Phi}.
\end{equation}
For $0 < u \ll 1$, we find
\begin{equation}
  \Phi(u) = 1/12 - u/30 + u^2/120 - u^3/105 + O(u^4)
  \label{eqn:Phizero}
\end{equation}
while for $u \gg 1$ we have $\Phi(u) \sim u^{-2}$, with a crossover
near $u = XL = 1$ (inset of Figure\ \ref{polfig}). The exponent of
$-2$ in the scaling regime is typically denoted by $-\gamma + 1$
\cite{DSFisher85,NarayanMiddleton94}, so that $\gamma = 3$.

The crossover and the scaling exponent can be motivated via record
sequences.  Given a current record $X$, the next record will occur on
average after $1/X$ sites.  Since all sites within this range are
forced to jump, this defines the correlation length $\xi \sim
X^{-\nu}$, with exponent $\nu = 1$.  The crossover to the saturated
regime occurs when $\xi$ is comparable to $L$, namely $u = XL \sim
L/\xi \sim 1$.  The size \eqref{eqn:SigmajLR} of an avalanche depends
on the product of its left and right extents and thus scales as
$X^{-2}$.

In the scaling regime, corresponding to large values of $u$, an
asymptotic expansion of $p_u(s)$ in $u$ shows that the distribution
obtains a scaling form in terms of the variable $a = u^2 s$,
\begin{equation}
  p(a) = 2 {\rm K_0}(2\sqrt{a}),
  \label{eqn:paK0}
\end{equation}
where ${\rm K_0}$ is the modified Bessel function, which decays at
large values of its argument as $e^{-2u\sqrt{s}}/(2u\sqrt{s})^{1/2}$.
For the $u$-values shown in Fig.~\ref{polfig}, the asymptotic form
\eqref{eqn:paK0} is indistinguishable from the exact result
\eqref{eqn:pdist}, explaining the collapse of the data.  The form of
the scaling variable $a$ can be understood by noting that $a = u^2 s =
X^2 \Sigma = \Sigma/\xi^2$; thus the avalanche sizes are measured in
units of $\xi^2$.

\section{Comparison with the full model}

\begin{figure}[t!]
  \begin{center}
  \end{center}
  \includegraphics[scale=0.3]{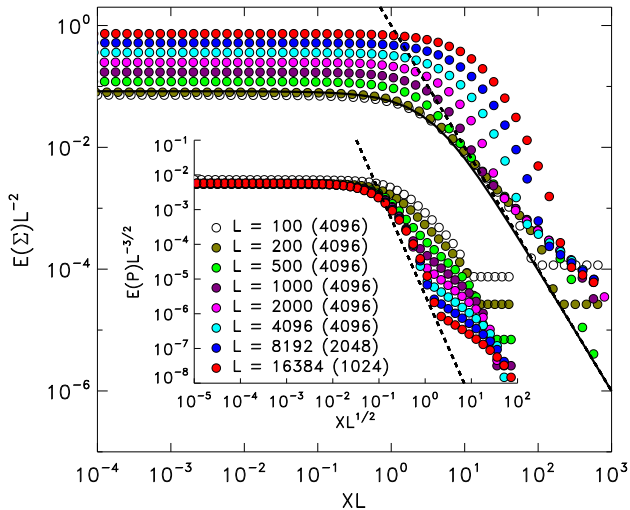}
  \caption[]{Finite-size scaling behavior of CDW model for $\lambda =
    10$. The system sizes and number of realizations are as indicated 
    in the legend of the inset.  Main
    figure: Finite-size scaling of the expected avalanche size at
    reduced force $X$, using the scaling of the toy model $\gamma =
    3$, $\nu = 1$. The solid line is the theoretical expression (22)
    obtained for the toy model, while the slanted dashed line
    indicates a power law with exponent $-2$. Inset: plot of the
    expected polarization $P = \Sigma/L$ against $X$ using the scaling
    predicted by Narayan {\em et al.}, $\gamma = 4$, $\nu = 2$. The
    dashed line indicates a power law with exponent $-3$.
}
  \label{th2tPolAve}
\end{figure}

The scaling exponents for the threshold-to-threshold evolution of the
toy model $(\nu = 1, \gamma=3)$ differ from those obtained via the
$4-\epsilon$ expansion of Narayan \emph{et al}
\cite{NarayanMiddleton94, NarayanFisher92}, $(\nu = 2,\gamma = 4)$ in
1d, but agree with those of the $d = 1$ CDW automaton model of Myers
and Sethna \cite{MyersSethna93}.  To investigate this discrepancy, we
have simulated the full CDW model \eqref{eqn:tildey} at small values
of $\lambda$. The results are shown in Fig.~\ref{th2tPolAve}.
Remarkably, our results suggest a crossover in the finite-size scaling
behavior, from that of the toy model at small system sizes $L$ to the
prediction of Narayan \emph{et al} at larger $L$.  For $\lambda = 10$
this crossover occurs around $L_c = 500$. Also, $L_c$
increases with $\lambda$. 

The presence of a $\lambda$-dependent length scale $L_c$ might also
explain why clean critical behavior in 1d CDW models was not seen in
numerical simulations carried out in the 1990s \cite{Middleton93,
  MyersSethna93}: the feasible system sizes of those years were
probably not large enough to push the simulations out of the crossover
region, ({\em see} Figs.~5 and 18 in\cite{Middleton93}).

We also performed numerical simulations of the toy model with flat initial
condition and find scaling of the polarization
that \emph{does} match that of Narayan \emph{et al} with $\gamma = 4$
and $\nu = 2$, which is also consistent with our analytical result, $P
\sim L^{3/2}$ at threshold, \eqref{eqn:flatscaling}.

\section{Discussion} 
We have presented an exactly solvable CDW toy model in 1d with a
critical depinning transition, exposing the roles played by the
disorder and the boundary conditions. The evolution towards threshold
is a process of breaking lower records on the coordinates of the
particles in their unit cells. This is a direct consequence of the
fact that as threshold is approached, larger segments are displaced
and the increasing stress at their boundaries has to be relieved by
particles which have sufficient room to advance without jumping.

Our numerical results for the scaling behavior of the full model
\eqref{eqn:tildey} show a crossover from the finite-size scaling of
the toy model for $L < L_c$ to the scaling predicted by Narayan {\em
  et al.}~at system sizes $L > L_c$. Moreover, the crossover length
scale $L_c$ increases with the pinning strength $\lambda$ so that the
critical behavior of the toy model can be obtained asymptotically at
large system sizes.
 
However, our toy model exhibits the critical behavior
of Narayan \emph{et al.}~if the initial condition is flat, $m_i = 0$
for all $i$.  Whereas the $(\pm)$-threshold states of the toy model
have well coordinates which differ in at most four locations, the flat
and $(+)$-threshold configurations are substantially different, and
the evolution in the rank representation is more complicated.  Instead
of a single trapezoidal bump growing by avalanches, we have separated
regions of activity which grow and merge.  We will report elsewhere
\cite{KM12} on a more detailed investigation of this behavior.

\acknowledgments DK and MM thank F.\ Rezakhanlou for
stimulating discussions. MM acknowledges discussions with H.J.\
Jensen and M.M.\ Terzi. He thanks the Berkeley Math department for
their kind hospitality during his sabbatical stay. 
This work was supported in part by by NSF grant DMS-1106526 and 
Bo{\u g}azi{\c c}i University grant 12B03P4.

\bibliography{km_short_arxiv}% Produces the bibliography via BibTeX.

\end{document}